\documentclass[aps,prb,twocolumn,showpacs]{revtex4}
\usepackage{graphicx}

\begin{document}

\title{Quantum interference structures in the conductance plateaus of gold
nanojunctions}

\author{A.~Halbritter, Sz.~Csonka, and G.~Mih\'aly}
\affiliation{Electron Transport Research Group of the Hungarian
Academy of Science and Department of Physics, Budapest University
of Technology and Economics, 1111 Budapest, Hungary}

\author{O.I.~Shklyarevskii$^\dag$, S.~Speller, and H.~van~Kempen}
\affiliation{NSRIM, University of Nijmegen, Toernooiveld 1, 6525
ED Nijmegen, the Netherlands}

\date{\today}

\begin{abstract}
The conductance of breaking metallic nanojunctions shows plateaus
alternated with sudden jumps, corresponding to the stretching of
stable atomic configurations and atomic rearrangements,
respectively. We investigate the structure of the conductance
plateaus both by measuring the voltage dependence of the plateaus'
slope on individual junctions and by a detailed statistical
analysis on a large amount of contacts. Though the atomic
discreteness of the junction plays a fundamental role in the
evolution of the conductance, we find that the fine structure of
the conductance plateaus is determined by quantum interference
phenomenon to a great extent.
\end{abstract}

\pacs{73.63.Rt, 73.23.Ad, 72.10.Fk, 72.15.Lh}

\maketitle

\begin{figure}
\includegraphics[width=\columnwidth]{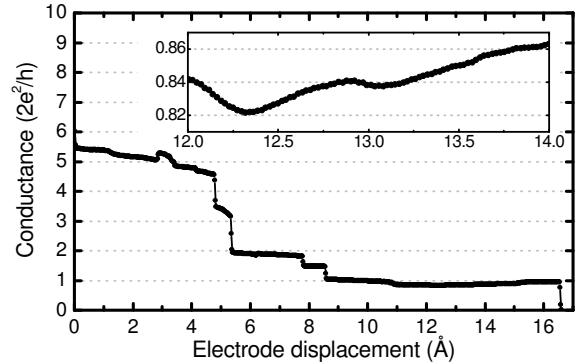}
\caption{\it Representative conductance trace recorded during the
break of a gold nanojunction. The inset shows a segment of the
last conductance plateau demonstrating the fine structure of the
conductance traces.} \label{condtrace.fig}
\end{figure}

The investigation of the mechanical and electrical properties of
atomic-sized metallic junctions has recently become an interesting
topic of nanoscience (for a review see
Ref.~\onlinecite{Agrait2003}). A contact with a single atom in the
cross section can be created by pulling a nanowire with a scanning
tunneling microscope (STM) or the mechanically controllable break
junction (MCBJ) technique. In such nanocontacts the coherent
quantum phenomena always interplay with the atomic granularity of
matter, as the wavelength of the electrons and the interatomic
distance are in the same order of magnitude. The atomic nature of
the junction is clearly demonstrated by the evolution of the
conductance during the break of the contact showing plateaus
alternated with sudden jumps (Fig.~\ref{condtrace.fig}). Force
measurements have shown that the conductance plateaus correspond
to the stretching of stable atomic configurations, whereas the
conductance jumps are related to atomic
rearrangements.\cite{Rubio1996} On the other hand, the statistical
analysis of a large amount of conductance vs.\ electrode
separation traces has shown signs of conductance quantization in
metals with loosely bound $s$ electrons.\cite{Krans1995} The
quantum nature of conductance is also reflected by the quantum
interference (QI) phenomenon of the electron waves scattered on
nearby impurities, which was reported in
Refs.~\onlinecite{Ludoph1999,Untiedt2000}. These works
investigated the interference patterns in the voltage dependence
of the conductance. In this paper we demonstrate that QI has a
definite influence on the structure of the conductance plateaus as
well, which arises due to the spatial variation of the electron
paths during the stretching of the junction.

The measurements were performed on high purity gold samples at
liquid Helium temperature with the MCBJ
technique.\cite{Agrait2003} The conductance histogram of Au shows
a sharp peak at the quantum conductance unit, G$_0=2e^2/h$. This
peak arises from the frequent occurrence of plateaus that are
accurately positioned at $1$\,G$_0$, as shown in
Fig.~\ref{condtrace.fig}. It was found, that these plateaus are
related to the conductance through a single gold
atom,\cite{Agrait2003} or through a chain of gold atoms in a
row.\cite{Yanson1998} In both cases the contact has a single
conductance channel with almost perfect
transmission.\cite{Scheer1998,Rubio2003} Theoretical studies have
pointed out that in gold the conductance of a monoatomic contact
is not sensitive to the amount of stretching, which could explain
the flatness of the last conductance plateau.\cite{Cuevas1998b} In
the experiments, however, the conductance plateaus always show a
fine structure, which are different during each rupture (for
examples see Ref.~\onlinecite{Untiedt2000} and the inset in
Fig.~\ref{condtrace.fig}). This feature could be naturally
explained by the atomic discreteness of the junction: as the
electrodes are pulled apart the overlap between the central atoms
changes, which alters the conductance of the contact. In this
paper we show that this interpretation is not satisfactory, and
the fine structure of the conductance plateaus is strongly
affected by quantum interference phenomenon.

\begin{figure}
\includegraphics[width=\columnwidth]{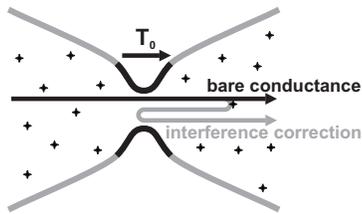}
\caption{\it Illustration for the quantum interference effect in
nanojunctions, following the model in
Ref.~\onlinecite{Ludoph1999}.} \label{interference.fig}
\end{figure}

The basic idea behind quantum interference in atomic-sized
junctions is illustrated in Fig.~\ref{interference.fig}. The
narrow neighborhood of the contact center can be considered as a
ballistic region with a transmission probability, $T_0$. The
electron wave that has travelled through the contact can be
partially reflected by impurities or lattice defects farther away
in the diffusive electrodes. This reflected wave goes back to the
contact, and a part of it is reflected back again by the contact
itself. This part of the wave interferes with the direct wave,
modifying the conductance of the junction. The net transmission
including the interference corrections can be written as:
\begin{eqnarray}\label{interference1.eq}
&&T(z,V)=\\
\nonumber &&=T_0(z)\left[1+\sum_j A_j
\cos\left\{\left(k_F+\frac{eV}{\hbar
v_F}\right)L_j+\Phi_j\right\}\right].
\end{eqnarray}
The total transmission is a function of the electrode separation,
$z$ and the bias voltage, $V$. The bare transmission of the
contact, $T_0$ is controlled  by the shape of the junction and the
overlap between the atomic orbitals, and accordingly it is
dependent on the electrode separation, $z$. It was shown that in
the voltage scale of the measurement the voltage dependence of
$T_0$ can be neglected.\cite{Nielsen2002,Brandbyge2001} In the
interference correction the sum runs over the various electron
trajectories; $L_j$ and $\Phi_j$ are respectively the path length
and the phase shift on a trajectory; and $k_F$ is the Fermi wave
number. The amplitude $A_j$ is determined by the scattering cross
section of the defect, the length of the path, and the reflection
of the contact. The differential conductance of the system is
obtained from the transmission as $G(z,V)=$G$_0\cdot T(z,V)$. (For
the sake of simplicity, a single conductance channel is
considered. The argumentation would be similar for multiple
channels as well.)

\begin{figure}
\includegraphics[width=\columnwidth]{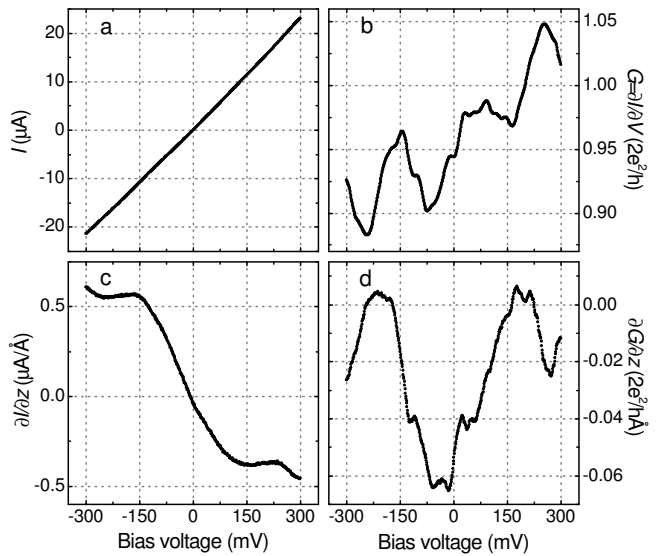}
\caption{\it The $I(V)$ curve (a), the $G(V)=\partial I/\partial
V$ curve (b), the $\partial I(V)/\partial z$ curve (c), and the
$\partial G(V)/\partial z$ curve (d) recorded on the same
single-atom gold junction.} \label{dIdz.fig}
\end{figure}

Quantum interference results in fluctuations in the conductance
when the interference conditions are tuned experimentally. If the
wave number of the electrons is changed by the bias voltage, QI
shows up as a small, random oscillation in the $G(V)$
curve.\cite{Ludoph1999,Untiedt2000} The interference pattern can
also be changed by tuning the phase factor of the electron paths
with magnetic field. In atomic-sized contacts, however, a magnetic
field of $\gtrsim 60$\,T would be required to have a considerable
influence on the interference, while a field of $1$\,T already
causes changes in the atomic arrangement of the contact due to
magnetostriction effects.\cite{Ludoph1999b} Here, we focus our
attention on quantum interference due to the variation of the {\it
length of the electron paths}. In nanojunctions the path length
naturally changes with the separation of the electrodes. To have a
complete period in the interference pattern the electrode
separation should be changed by one wavelength of the electrons.
Experimentally, such a displacement is not possible without a
jump-like atomic rearrangement, which abruptly changes the
interference pattern. From this reason, only shorter parts of the
conductance plateaus can be studied, like that in the inset of
Fig.~\ref{condtrace.fig}. The fine structure of these short
segments can originate both from the QI phenomenon and from the
electrode separation dependence of the bare transmission,
$T_0(z)$. In the following we show experimental techniques, that
can tell ``to what extent these two phenomena are involved in the
evolution of the plateaus''. To investigate the fine structure of
the conductance traces, we have studied the local slope of the
plateaus by two different methods.

The first approach examines the effect of bias voltage on the
plateaus' slope on individual junctions. Figure \ref{dIdz.fig}
shows the current (panel a), and the derivative of the current
with respect to the electrode separation (panel c) recorded as a
function of the bias voltage. The two curves were measured
simultaneously on the same junction. The electrode separation was
modulated by applying a sine-wave voltage on the piezo element.
The oscillation of the separation had a typical amplitude of
$0.1$\,\AA. As the bias voltage was varied, the current was
detected both by a current meter measuring the DC component and a
lock-in amplifier recording the response to the modulation. The
signal of the current meter provided the $I(V)$ curve, whereas the
lock-in measured the value of $\partial I/\partial z$. The
differential conductance, $G(V)$ and the slope of the plateau,
$\partial G/\partial z$ was determined by numerical
differentiation (Fig.~\ref{dIdz.fig}b and \ref{dIdz.fig}d,
respectively). These curves are reproducible to the very small
details as long as the same contact is measured. When the junction
is changed a completely new structures appear in the curves, as
expected from QI phenomenon.

Assume that the dependence of the differential conductance on the
electrode separation, $z$ is attributed solely to the bare
transmission $T_0(z)$. In this case the slope of the conductance
plateau can be written as:
\begin{equation}
\frac{\partial G(z,V)}{\partial z}=\frac{1}{T_0(z)}\frac{\partial
T_0(z)}{\partial z}G(z,V),
\end{equation}
i.e.\ the voltage dependence of $\partial G/\partial z$ is simply
proportional to $G(V)$. This, however is disproved by the
experimental results shown above. The oscillatory patterns of the
$G(V)$ curve and the $\partial G(V)/\partial z$ curve in
Fig.~\ref{dIdz.fig} do not coincide. Furthermore, in the $G(V)$
curve the oscillations have a typical amplitude of $10\%$ compared
to the mean value of $G=0.96$\,G$_0$, while in the $\partial
G(V)/\partial z$ curve the relative amplitude of the oscillations
is more than 10 times larger.

These observations can only be explained, if the change of the
path lengths $L_j \rightarrow L_j+dz$ is also taken into account
as the electrode separation is varied by $dz$. Then, the
derivative of the transmission with respect to $z$ is written
as:\cite{note2}
\begin{eqnarray}\label{interference3.eq}
&&\frac{\partial T(z,V)}{\partial z}\simeq\frac{\partial
T_0(z)}{\partial z}-\\
\nonumber &&-T_0(z)\sum_j k_F A_j\sin
\left\{\left(k_F+\frac{eV}{\hbar v_F}\right)L_j+\Phi_j\right\}.
\end{eqnarray}
Based on this formula, $\partial T_0/\partial z$ is well
approximated with the mean value of the $\partial G(V)/\partial z$
curve, which is $\simeq -0.023$\,\AA$^{-1}$. The amplitude of the
interference correction is characterized by the standard
deviation: $\simeq 0.022$\,\AA$^{-1}$. It shows, that the
variation of the plateau's slope due to QI is comparable to the
separation dependence of the bare transmission. The comparison of
the formulas (\ref{interference1.eq}) and (\ref{interference3.eq})
shows, that the amplitude of the oscillatory term changes by a
factor of $k_F$, while the constant term changes by $(\partial
T_0/\partial z)/T_0$ due to the differentiation. According to
measurements on several contacts, $\partial T_0/\partial z$ is
typically below $0.05$\,\AA$^{-1}$, which is smaller by an order
of a magnitude than $k_F\simeq0.6$\,\AA$^{-1}$. This explains that
the contribution of QI is highly enhanced in the $\partial
G/\partial z$ curves, while in the $G(V)$ curve it only gives a
minor correction.

The above measurements were performed on individual contacts. In
the following we present a second approach, investigating the
statistical properties of the slope of the conductance plateaus.
Independent atomic configurations with different set of the
interference parameters ($A_j$, $L_j$ and $\Phi_j$) can be
naturally created by repeating the break of the junction several
times. The data set for the statistical analysis was obtained by
recording $\sim 15000$ independent conductance vs.\ electrode
separation traces at fixed bias voltage. The typical acquisition
rate was $50$\,points/\AA. The slope of the plateaus was
determined by numerical differentiation. The derivative was
calculated at each point of the conductance plateaus, however the
jump-like changes between two plateaus -- corresponding to sudden
atomic rearrangements -- were excluded from the analysis.

In the mean value of $\partial G/\partial z$ the interference
corrections cancel out due to their random distribution around
zero, thus the average slope of the plateaus is only determined by
the bare transmission:
\begin{equation}\label{stat1.eq}
\left\langle \frac{\partial G}{\partial z}\right\rangle=G_0
\left\langle \frac{\partial T_0}{\partial z}\right\rangle.
\end{equation}
The proper quantity to study QI is rather the mean square
deviation of $\partial G/\partial z$, which contains the
interference term beside the properties of the bare contact (see
Eq.~\ref{interference3.eq}):
\begin{equation}\label{stat2.eq}
\frac{\sigma^2_{\partial G/\partial z}}{G^2_0}=\sigma^2_{\partial
T_0/\partial z}+\underbrace{\frac{1}{2}T^2_0k^2_F\sum_j\langle
A^2_j \rangle}_{\textstyle \sigma^2_{QI}}.
\end{equation}
The squared amplitude, $A_j^2$ is proportional to the probability
that an electron is reflected back by the contact, $R_0=1-T_0$.
Therefore, the interference term in the mean square deviation
vanishes both at $T_0=1$ and $T_0=0$.

\begin{figure}[t!]
\includegraphics[width=\columnwidth]{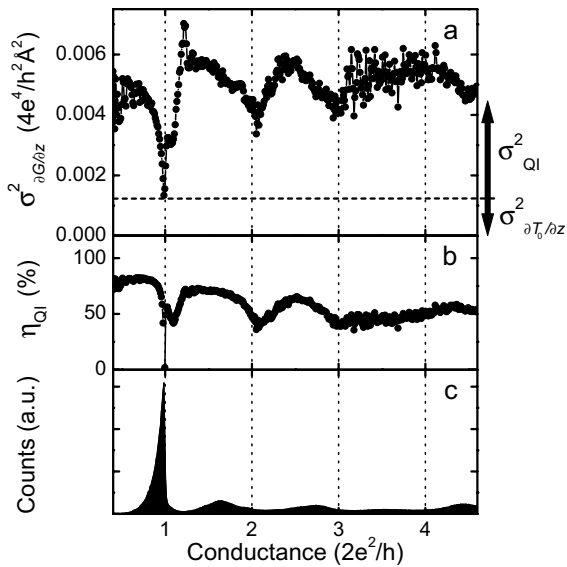}
\caption{\it Panel (a) shows the standard deviation of the plateau
slope, $\partial G/\partial z$ as the function of the conductance.
The arrows indicate the separation of the QI term from
$\sigma^2_{\partial T_0/\partial z}$. Panel (b) presents the
relative contribution of QI to the plateaus' slope. In panel (c)
the conductance histogram is presented for the same data set.}
\label{fluct.fig}
\end{figure}

Gold junctions with a few atoms ($\le4$) in the cross section show
the saturation of the channel transmissions, which means that a
new channel only starts to open, if the previous ones are almost
completely open. Due to this behavior at the quantized conductance
values all transmission probabilities are close to unity or zero,
thus the quantum interference is suppressed.\cite{Ludoph1999} If
QI gives a detectable contribution to the slope of the plateaus,
the $\sigma^2_{\partial G/\partial z}(G)$ curves should also
exhibit the quantum suppression at the multiples of G$_0$. This
phenomenon is clearly resolved in our experiments: the mean square
deviation of the plateaus' slope exhibit pronounced minima
accurately placed at $1, 2,$ and $3$\,G$_0$
(Fig.~\ref{fluct.fig}a). In contrast, the second and the third
peak in the conductance histogram are significantly shifted from
the integer values (Fig.~\ref{fluct.fig}c). It demonstrates that
the minima in $\sigma^2_{\partial G/\partial z}$ are a consequence
of a pure quantum phenomenon, and they are not related to the
preferred atomic configurations shown by the peaks in the
histogram.

The suppression of QI at the quantized values gives a possibility
to estimate the contribution of the quantum interference term to
the slope of the plateaus. According to
Ref.~\onlinecite{Ludoph1999} the magnitude of the quantum
suppression is almost $100\%$ at $1$\,G$_0$, while at higher
quantized values it is decreasing. Therefore, we attribute the
nonzero minimum value of $\sigma^2_{\partial G/\partial z}$ at
$1$\,G$_0$ purely to the scattering of the bare properties,
$\sigma^2_{\partial T_0/\partial z}$. The interference term in
Eq.~\ref{stat2.eq}, $\sigma^2_{QI}$ is approximated by subtracting
$\sigma^2_{\partial T_0/\partial z}$, which is considered as a
constant background.\cite{note3} The relative amplitude of QI in
the slope of the plateaus can be characterized by the quantity
$\eta^{\ }_{QI}= G_0\cdot\sigma^{\ }_{QI}/\sqrt{\langle\partial
G/\partial z\rangle^2+\sigma^2_{\partial G/\partial z}}$. This
curve takes values larger than $50\%$ (Fig.~\ref{fluct.fig}b),
which demonstrates that the influence of QI on the slope of the
plateaus is dominating over the features due to the atomic
arrangement of the bare contact.

For a more quantitative description of the observations we have
performed a calculation following the model in
Ref.~\onlinecite{Ludoph1999}. In the $V\rightarrow 0$ limit the
standard deviation of the plateau's slope due to the QI terms can
be written as:
\begin{equation}
\sigma^2_{QI}=\frac{24}{\sqrt{\pi}(1-\cos\gamma)}\frac{1}{l_e^{2}}\cdot
\sum_{n=1}^{N}T_n^2(1-T_n)\\
\end{equation}
This formula already treats a multichannel situation, where $T_n$
is the transmission of the $n$-th channel, $\gamma$ is the opening
angle of the contact, and $l_e$ is the elastic mean free path of
the electrons. From the measured amplitude of $\sigma^{2}_{QI}$
the elastic mean free path is estimated as $\sim 5$\,nm, which is
in good agreement with previous results.\cite{Ludoph1999}

Concluding,  we have investigated the structure of the conductance
plateaus in gold nanocontacts. We have studied the voltage
dependence of the slope of the conductance plateaus on individual
junctions. The $\partial G(V)/\partial z$ curves have shown a
strong oscillatory deviation from the mean value, which is an
order of a magnitude larger than the conductance fluctuations in
the G(V) characteristics. This feature could only be described by
quantum interference due to the spatial modulation of the
interference paths. In order to support these results we have
performed a statistical analysis of the plateaus' slope for a
large amount of junctions. The quantum suppression of $\sigma^{2
}_{\partial G/\partial z}$ at the quantized conductance values
have provided an even stronger proof for the significant presence
of QI. With our analysis the contributions of quantum interference
and the strain dependence of the local atomic configuration to the
plateaus' slope could be separated. The results have shown that
the quantum interference phenomenon and the atomic discreteness of
the junction have a similarly strong influence on the fine
structure of the conductance plateaus.

The authors acknowledge the financial support from the ``Stichting
FOM'' and the Hungarian research funds OTKA TS040878, T037451.


\begin{thebibliography}{13}
\bibitem[{not({\natexlab{a}})}]{note2}
\bibinfo{note}{In the experiments the applied voltage is much smaller than the
  Fermi energy, so the term $eV/\hbar v_F$ was neglected beside $k_F$. The term
  $\partial T_0/\partial z \sum_j A_j\cos(...)$ was also neglected being an
  order of a magnitude smaller than $\partial T_0/\partial z$.}
\end{thebibliography}
\end{document}